\documentclass[a4paper,11pt]{amsart}

\usepackage{enumerate}
\usepackage{amssymb,amsfonts}
\usepackage{exscale}

\newcommand{\be}{\begin{equation}}
\newcommand{\ee}{\end{equation}}
\newcommand{\bea}{\begin{eqnarray*}}
\newcommand{\eea}{\end{eqnarray*}}
\newcommand{\beq}{\begin{eqnarray}}
\newcommand{\eeq}{\end{eqnarray}}

\newcommand{\RR}{\mathbb{R}}

\newcommand{\NN}{\mathbb{N}}
\newcommand{\ZZ}{\mathbb{Z}}

\newcommand{\EE}{\, \mathbb{E} \,}
\newcommand{\PP}{\mathbb{P}}

\renewcommand{\r}{\right}
\renewcommand{\l}{\left}

\newcommand{\Lp}{\mathop{\Lambda^+}}

\renewcommand{\L}{\Lambda}
\renewcommand{\Lp}{\Lambda^+}

\newcommand{\diam}{\mathop{\mathrm{diam}}}
\newcommand{\vol}{\mathop{\mathrm{vol}}}    
\newcommand{\loc}{\mathop{\mathrm{loc}}}

\newcommand{\supp}{\mathop{\mathrm{supp}}}

\newcommand{\Tr}{\mathop{\mathrm{Tr}}}

\newtheorem{thm}{Theorem}[section]

\newtheorem{prp}[thm]{Proposition}

\theoremstyle{remark}

\newtheorem{bsp}[thm]{Example}

\begin{document}
\title[Density of states for some alloy type models with changing sign]{Existence of the density of states\\ for some alloy type models with single site potentials of changing  sign}

\author[I.~Veseli\'c]{Ivan Veseli\'c}
\address{Fakult\"at f\"ur Mathematik\\Ruhr-Universit\"at Bochum, Germany} 
\email{Ivan.Veselic@ruhr-uni-bochum.de}
\urladdr{www.ruhr-uni-bochum.de/mathphys/ivan}
 

\begin{abstract}
We study spectral properties of ergodic random Schr\"{o}dinger operators on $L^2 (\RR^d)$. The density of states is shown to exist for a certain class of alloy type potentials with single site potentials of changing sign. The Wegner estimate we prove implies Anderson localization under certain additional assumptions. For some examples we discuss briefly some properties of the common and conditional densities of the random coupling constants used in the proof of the Wegner estimate.\\[.5em]
{ \sc Sa\v zetak. }
Analiziraju se spektralna svojstva ergodi\v ckih slu\v cajnih
 Schr\"{o}\-din\-gerovih operatora na $L^2 (\RR^d)$. Dokazuje se, da
 gusto\'ca stanja postoji za odredjenu klasu potencijala tipa legure, kod
 kojih pojedina\v cni potencijal mijenja predznak. Uz odredjene dodatne
 uvjete Wegnerova ocjena koju dokazujemo  implicira fenomen Andersonove
 lokalizacije.  Na osnovu primjera  promatramo neka svojstva
 zajedni\v cke i uvjetne gusto\'ce slu\v cajnih konstanti veze. Te
 gusto\'ce se koriste u dokazu  Wegnerove ocjene.
\end{abstract}
\maketitle

\section{Alloy type model and the integrated density of states}

We consider Schr\"{o}dinger operators with a potential which is a stochastic process ergodic with respect to translations from $\ZZ^d$. Such operators  model  quantum mechanical Hamiltonians which govern the motion of single electrons in disordered solids. The spectral properties of the Schr\"{o}dinger operator are related to the dynamical behaviour of the electron wave packets and thus to the charge transport properties of the described solid, cf.~e.g.~\cite{Bonch-BruevichEEKMZ-1984,EfrosS-84,LifshitzGP-88}.

The random potential considered in this note is of {\em alloy} or {\em continuous Anderson} type:
\begin{eqnarray*}
V : \Omega \times \RR^d \to \RR ,  \quad V_\omega(x) = \sum_{k \in \ZZ^d } u(x-k).
\end{eqnarray*} 
The function $u\in L^p$ ($p=2$ for $d\le 3$ and $p > \frac{d}{2}$ for $d\ge 4$) whose translates generate $V_\omega$ is called {\em single site potential}. The "strength" of  the potentials $u(\cdot-k)$ associated to different lattice sites $k \in \ZZ^d$ is a random variable $\omega_k: \Omega \to \RR$. We assume that the sequence $(\omega_k )_{k \in\ZZ^d}$ is independent and  identically distributed (iid),  and  moreover that each $\omega_k$ is distributed according to the probability measure $\mu$ which has a density $f $ in the Sobolev space  $ W_c^{1,1}(\RR)$. We may identify the probability space $(\Omega, \PP)$ with the space $\times_{k\in \ZZ^d} \RR$ equipped with the product measure $\otimes_{k \in\ZZ^d} \mu$. 
Thus an element $\omega \in \Omega$ is a random vector $(\omega_k)_{k \in\ZZ^d}$. 

The {\em alloy type model} is the random Schr\"{o}dinger operator, or more precisely, the random family of Schr\"{o}dinger operators, given by
\begin{equation}
\label{Homega}
H_\omega := H_0  + V_\omega, \quad \omega \in \Omega, \qquad H_0 := -\Delta +V_0 
\end{equation}
where $V_0 \in L_{\loc}^p(\RR^d)$ is a $\ZZ^d$-periodic potential and $p$ is chosen depending on the dimension, as above. 

A large part of the mathematical literature and results on random Schr\"{o}dinger operators are devoted to the model just described, certain derivatives thereof, cf.~e.g.~\cite{Klopp-93,CombesHM-96,HundertmarkK-2000} and 
its discrete version on $l^2(\ZZ^d)$ --- actually the original model studied by Anderson \cite{Anderson-58}. An extensive list of references can be found in the textbooks \cite{CyconFKS-87b,Kirsch-89a,CarmonaL-1990,PasturF-1992,Stollmann-2001}. 

The name alloy type model stems from the fact that one can consider $u(\cdot-k)$ as an atomic potential, say of screened Coulomb type, due to a nucleus sitting at the lattice site $k \in \ZZ^d$. While the lattice structure of the solid is fixed and all atoms are assumed to generate a potential of the same {\em shape}, the distribution of the different types of atoms on the lattice sites is random. This is then modelled by random nuclear charges $\omega_k$ which appear as {\em coupling constants} in front of the atomic potential $u(\cdot-k)$ and determine its "strength".

Under the above assumptions on $H_\omega$ it is well known that the spectrum and its measure theoretic components are almost surely non-random. More precisely, there exist $\Omega' \subset \Omega$ of measure one and a subset of the reals $\Sigma $ such that
$
\sigma(H_\omega) = \Sigma $ for all $ \omega \in \Omega'.
$
The analog statement is true for the sets $\sigma_{disc} (H_\omega),\sigma_{ess} (H_\omega),\sigma_{ac} (H_\omega),\sigma_{sc} (H_\omega)$ and $\sigma_{pp} (H_\omega)$. (Here we denote by $\sigma_{pp} $ the {\em closure} of the set of eigenvalues.)
This property of the spectrum is called {\em selfaveraging}. It is due to the fact that on the infinite configuration space $\RR^d$ the fluctuations of the potential $V_\omega$, which are present in finite size cubes, cancel out thanks to the iid property of the random coupling constants.

Part of the spectral properties of the family of Schr\"{o}dinger operators $H_\omega$ is contained in the {\em integrated density of states} (IDS). It can be used to calculate the basic thermodynamic quantities of the corresponding non-interacting many particle system, cf.~e.g.~\cite{Huang-1987,Dorlas-1999}. In physical terms, the IDS counts the number of energy levels per unit volume, up to a given energy $E$.  The possible existence of continuous spectrum makes a careful mathematical definition of the IDS necessary, which is as follows: Denote by $\Lambda=\Lambda_l$ the cube $[0,l[^d$ and with $H_\omega^l$ the restriction of $H_\omega$ to the interior of $\Lambda_l$ with
periodic boundary conditions.  Then the normalized eigenvalue counting functions
\begin{equation}
N_\omega^l (E) = l^{-d} \# \{ i | \ \lambda_i (H_\omega^l) < E \}
= l^{-d} \Tr P_\omega^l(]-\infty,E[) 
\end{equation}
of $H_\omega^l$ converge for almost all $\omega$ to a limit
$N(E) := \lim_{ l \to \infty} N_\omega^l(E) $ which is $\omega$-independent. This convergence holds true for all $E$ which are continuity points of $N$. Here we denote with $P_\omega^l(I)$ the spectral projection of $H_\omega$ onto the interval $I$. 
Since the limit $N$ is independent of the randomness, we encountered another selfaveraging quantity. Its set of points of increase ${\mathcal I} := \{ E\in\RR| \, N(E+\epsilon) - N(E-\epsilon) > 0 \}$ coincides with the a.s.~spectrum $\Sigma$.

This note is organized as follows. In the next section we state our main theorem and related results of other authors as well as the implications for Anderson localization, Section 3 contains a sketch of the proof of the main theorem and the last section is devoted to the discussion of the proof of localization using the multiscale analysis and to Wegner estimates for alloy type potentials with dependent random coupling constants. Furthermore we discuss some technical differences of the use of the common  and the conditional density of the coupling constants. 

\section{Main theorem: A Wegner estimate for indefinite potentials}
A Wegner estimate \cite{Wegner-81} is a assertion about the regularity of the finite volume IDS $N_\omega^l$ which may imply the H\"{o}lder continuity of the IDS on $\RR^d$ or even the existence of its derivative  $d N/d E$, the {\em density of states} (DOS).
Note that in  the following result the single site potential may be indefinite, i.e.~take values of both signs.
\begin{thm}[\cite{Veselic-2000b,Veselic-2001}]
\label{theorem1}
Let $ L^p(\RR^d) \ni w \ge \kappa \chi_{[0,1]^d}$ with  $\kappa>0 $ and $p=2$ for $d\le 3$ and $p >d/2$ for $d\ge 4$.
Let $ \Gamma \subset \ZZ^d$ be finite, the {\em convolution vector} $\alpha=(\alpha_k)_{k \in\Gamma}$ satisfy 
$\alpha^* := \sum_{k \not= 0} | \alpha_k | < |\alpha_0|$, and
the single site potential be of {\em generalized step function} form:
\begin{equation}
\label{u}
u(x) = \sum_{k \in \Gamma} \alpha_k \ w(x -k). 
\end{equation}
Then there exists for all $E \in \RR$ a constant $C=C(E)$ such that
\begin{equation}
\label{resultat1}
\int_\Omega \l [ \Tr P_\omega^l( \l [E -\epsilon,E \r ]) \r ] \, d \PP(\omega)
\le C \  \epsilon \, l^d , \quad \forall \, \epsilon \ge 0.
\end{equation}
\end{thm}
The theorem implies that the DOS exists for a.e.~$E$ and is locally uniformly bounded: $d N(E)/d E \le C(E_1)  $ for all $E \le E_1$.  The following result with V.~Kostrykin applies to uniform densities. 
\begin{prp}[\cite{KostrykinV-2001}]
\label{theorem2}
The assertion of Theorem \ref{theorem1} holds true if $f$ is the uniform density on an interval and $\Gamma \subset \{k \in \ZZ^d | \, k_i \ge 0 \ \forall \ i = 1, \dots,d  \}$.
\end{prp}
First Wegner estimates for indefinite alloy type potentials were derived in \cite{Klopp-95a}.
In \cite{HislopK-2001} P.~Hislop and F.~Klopp combine the techniques from \cite{Klopp-95a} and  \cite{CombesHN-2001} to prove a Wegner estimate valid for general indefinite single site potentials and for  energy intervals at edges of $\sigma(H_\omega)$. They assume the single site potential $u$ not to vanish identically and to be in $C_c \cap l^1(L^p)$ with $p\ge \min(d,2)$. The density $f\in L_c^\infty$ of the random variable $\omega_0$ (which may be in fact the conditional density w.r.t.~$\omega^{\bot 0} := (\omega_k)_{k \not= 0}$) is assumed to be locally absolutely continuous. For any $\beta < 1$ and any energy interval $I$ below the spectrum of the unperturbed operator $H_0$ they prove
\begin{eqnarray*}
\PP \{\sigma (H_\omega^l) \cap I \not= \emptyset \} \le C \, |I|^\beta \, l^d
\end{eqnarray*}
where the constant $C$ depends only on $\beta, d$ and $dist(I, \sigma(H_0))$. With a sufficiently small global coupling constant $\lambda$ the same result holds for the operator $H_0 + \lambda V_\omega$ for $I$ in an internal spectral gap of $H_0$. The results of \cite{HislopK-2001} extend to more general models including certain operators with random magnetic field.

The literature on Wegner estimates for multidimensional alloy type models includes \cite{KotaniS-87,CombesH-94b,BarbarouxCH-1997,KirschSS-1998a,KirschSS-1998b,Stollmann-2000b,CombesHN-2001,HupferLMW-2001,KirschV-2001}.

Theorem \ref{theorem1} and Proposition \ref{theorem2} imply a localization result if the negative part $u_-$ of the single site potential is sufficiently small.

\begin{thm}[\cite{Veselic-2000b}]
\label{loc}
Let $H_\omega$ satisfy the assumptions of Theorem \ref{theorem1} or Proposition \ref{theorem2}, let $w$ have compact support and $E$ be a boundary point of $\sigma(H_\omega)$. Let furthermore either
\begin{enumerate} 
\item
 $V_0$ be symmetric w.r.t.~refelections along the coordinate axes and $E=\inf\sigma(H_\omega)$, or
\item
$\supp f = [\omega_-,\omega_+]$ and for $\tau > d/2$ and   $h\ge 0$ sufficiently small let the density $f$ satisfy 
$ \int_{\omega_-}^h f  \le h^\tau $ and $ \int_{\omega_+ -h}^{\omega_+} f \le h^\tau$.
\end{enumerate} 
Then there exist $\epsilon,r >0$ such that for $\sum_{\alpha_l < 0} \alpha_l \ge -\epsilon $     
\begin{equation}
    [E-r,E+r] \cap \sigma_{c} ( H_\omega ) = \emptyset 
    \     \mbox{ and } \ \sigma (H_\omega) \cap [E-r, E+r]   \not=\emptyset 
    \end{equation}     
The eigenfunctions of $H_\omega$ with eigenvalues in $[E-r, E+r]$ decay exponentially.
\end{thm}
In \cite{HislopK-2001} this result has been generalized to a larger class of single site potentials using results from 
\cite{Klopp-1999} and an abstract version of the smallness of $u_-$.
In one dimension the alloy type model has pure point at all energies regardless of the sign properties of $u$ \cite{Stolz-2000}. The proof does not use a Wegner estimate.

\section{Sketch of the proof of Theorem \ref{theorem1}}
For simplicity we assume $w= \chi_{[0,1]^d}$. Let $\tilde{\Lambda} := \Lambda \cap \ZZ^d$, $\EE$ denote the expectation w.r.t.~$\PP$ and $I := [ E_1, E_2[$ an energy interval. Thus we have
$ 
\EE \l [ N_\omega^l (E_2) - N_\omega^l (E_1  ) \r ]
=  l^{-d} \EE \l [ \Tr P_\omega^l(I) \r ].
$
We abbreviate $\chi_j:= \chi_{[0,1]^d+j}$ and  infer from \cite{CombesH-94b}  the inequality
\begin{equation}
\label{CH}
\EE \l [ \Tr P_\omega^l(I) \r ]
\le e^{E_2} C_V \sum_{j \in \tilde{\Lambda}} 
\, \l \|  \EE \l [ \chi_j P_\omega^l (I) \chi_j \r ] \r \|
\end{equation}
which corresponds to a partition of unity. Here the constant $C_V$ depends only on $V_0,u$ and $f$.
To estimate $ \EE [ \langle \phi , \chi_j P_\omega^l(I) \chi_j
\phi \rangle ]$ for any normalized  $\phi \in L^2 (\Lambda_l)$ we introduce a transformation of coordinates on the probability space $\Omega$.

For each cube $\L= \L_l$ denote $\Lambda^+ :=\{ \lambda - \gamma | 
\ \lambda \in \tilde{\Lambda}, \gamma \in \Gamma \}$. The operator $H_\omega^l$ depends only on the truncated random vector $(\omega_k)_{k \in\Lp}\in \RR^{\# \Lp}$.  On such vectors acts  a block Toeplitz transformation $A_\L := \{ \alpha_{j-k} \}_{j, k \in \Lp} $ induced by the convolution vector $\alpha$. The transformation has an inverse $B_\L= \{b_{k,j}\}_{k,j \in \Lp}= A_\L^{-1}$ which is bounded  in the row-sum norm $ \| B_\L \| \le \frac{1}{1-\alpha^*}$. We drop now the subscript $\L$ and denote with $\eta := A \omega$ the vector of the transformed random coordinates. They have the common density 
\begin{equation}
\label{k}
 k( \eta) =  k_\Lambda ( \eta) = | \det B| \, F(A^{-1}\eta)
 \end{equation}
 where $ 
F(\omega)=\prod_{k \in \Lambda^+} f \l ( \omega_k \r)$ is the original density of the $\omega_k$.
We calculate the potential $V_\omega$ written as a function of $\eta$ (and $x \in \L$): 
$$
V_\omega (x)  = V_{B\eta}(x)=
\sum_{j \in \tilde{\Lambda}} \eta_j \chi_j (x).
$$
In the new representation of the potential the single site potentials are non-negative, so we can to use a spectral averaging formula \cite{CombesH-94b} 
 \begin{equation}
 \label{spav}
  \int_{\RR} d\eta_j \, k(\eta ) \, s(\eta )
 \le 
 |I| \ \sup_{\eta_j} | k (\eta)|, \quad \mbox{ where }s(\eta ) := \langle \phi, \chi_j P_{B\eta}^l (I) \chi_j \phi \rangle.
 \end{equation}
Denote $L= \# \Lambda^+$. Fubini, (\ref{spav}) and the fundamental theorem of calculus   give
\begin{equation}
\label{sup}
\int_{\RR^L} d\eta \, k( \eta ) \, s( \eta ) 
\le |I| \, \int_{\RR^{L-1}} d\eta^{\bot j} \,  
 \sup_{\eta_j} | k (\eta)|
   \le 
|I| \, \int_{\RR^{L}} d\eta \       | (\partial_j k)(\eta) | 
.
\end{equation} 
The last integral equals $|\det A| \, \int_{\RR^{L}} d\omega \,   | (\partial_j k ) (A \omega  ) |$ which is bounded by \\
$\| f' \|_{L^1} \sum_{k \in \Lambda^+} |b_{k,j}|$. The proof of the theorem is finished by the  estimate 
\begin{equation}
\label{main-estimate}
\EE \l [ \langle \phi , \chi_j P_\omega^l(I) \chi_j  \phi \rangle \r ] 
\le 
|I| \ \| f' \|_{L^1} \|B\|.
\end{equation}

\section{Discussion and Applications}
\label{discuss}
\subsection{Localization}
The main application of Wegner estimates --- apart from establishing the regularity of the IDS --- is the proof of localization, i.e.~the existence of dense pure point spectrum 
of $H_\omega$ in certain energy regions. For  multidimensional Schr\"{o}dinger operators on $L^d(\RR^d)$ the multiscale analysis of J.~Fr\"{o}hlich and T.~Spencer \cite{FroehlichS-83} provides a method for proving localization by induction over cubes in $\RR^d$ of larger and larger side lengths. To start the multiscale induction one has to know certain off-diagonal decay estimates of the kernel of the resolvent of $H_\omega^l$ at an initial length scale. For non-negative $u$ there are rather well understood sufficient conditions for this estimates. The Wegner estimate is needed, too, as an a priori estimate for the multiscale analysis, namely to control the induction step when going from one scale to another, larger one. 

We sketch the main estimates in the multiscale analysis under the assumptions of Theorem \ref{loc}. 
  For the following it is convenient to slightly change the notation and denote by $\Lambda=\Lambda_l(x)=[-l/2,l/2]^d+x$ the cube of side length $l$ centered at $x$. The characteristic function of $\Lambda_l(x)-\Lambda_{l-2}(x)$, resp.~$\Lambda_{l/3}(x)$, is abbreviated as $\chi^+$, resp.~$\chi^-$. The cube $\Lambda$ is called {\em $(\gamma,E)$-good} if the associated resolvent satisfies the following off-diagonal decay estimate in operator norm
\begin{eqnarray*}
\| \chi^+ (H_\omega^\Lambda -E)^{-1} \chi^- \|   \le   e^{-\gamma \, l}.
\end{eqnarray*}
The initial scale estimate for the multiscale analysis  is satisfied if 
\begin{equation}
\label{ini}
\PP \{ \forall E \in I : \Lambda_l(x) \mbox{ is $(\gamma_0,E)$-good }  \} \ge 1-l_0^{-\xi_0}
\end{equation}
holds for $x\in \ZZ^d, l_0\in \NN$ sufficiently large and $\gamma_0 \ge l_0^{\beta-1}, \beta, \xi_0 > 0$. See \cite{Veselic-2000b} how it is derived in the situation of  Theorem \ref{loc}. It serves as the induction anchor. The geometric resolvent equation for two cubes $\Lambda\subset \Lambda', \phi\in C_c^1(\Lambda)$ and $z \in \rho (H_\omega^\Lambda) \cap \rho (H_\omega^{\Lambda'})$
\begin{eqnarray*}
(H^\Lambda -z)^{-1} \phi = \phi (H^{\Lambda'} -z)^{-1} + (H^\Lambda -z)^{-1}  [(\nabla \phi)\nabla + \Delta\phi] (H^{\Lambda'} -z)^{-1} 
\end{eqnarray*} 
and (some weak form of) a Wegner estimate 
$$
\PP\{\sigma (H_\omega^\Lambda) \cap I \not= \emptyset  \} \le C\, |I|^a \,\vol(\Lambda)^b 
$$
with $a\in ]0,1], b\in [1,\infty[$ imply for $x,y\in \ZZ^d, d(x,y)\ge \diam \supp u$
\begin{eqnarray*}
\PP \{ \forall E \in I : \Lambda_l(x) \mbox{ or $\Lambda_l(x) $ is $(\gamma,E)$-good }  \} \ge 1-l^{-2\xi}
\end{eqnarray*}
for some $\alpha\in ]1,2[,\gamma, \xi>0$ and  all $l_k ,k\in \NN$, where  $l_{k-1}  ^\alpha \le l_{k}  \le l_{k-1}^\alpha +6,k\in \NN$. 
This establishes the exponential decay of the resolvent on arbitrary large scales with high probability. Using a priori estimates on the (polynomial) growth of eigenfunctions \cite{Berezanskii-68} one concludes that they actually  decay exponentially. 

Details on the multiscale analysis can be found, e.g.,  in \cite{Stollmann-2001}
and some recent developments in \cite{GerminetK-2001a,GerminetK-2001b}.

\subsection{Dependent random coupling constants}
The proof of Theorem \ref{theorem1} uses the transformation of $V_\omega$ into an alloy type potential with non-negative single site potential $w\in L^p(\RR^d)$ and (negatively) correlated coupling constants $(\eta_k)_{k\in\ZZ^d}$. Thus we can interpret it as an Wegner estimate for the Schr\"{o}dinger operator $H_\eta= -\Delta+ V_0 + \sum_{k \in \ZZ^d}\eta_k w(\cdot-k)$. This implies the following Wegner estimate for dependent coupling constants:
\begin{prp}
Let $w\ge \kappa \chi_{[0,1]}, \kappa >0$ and for each $l\in\NN$ the common density $k_l$ of $\eta^l=(\eta_k)_{k \in \Lambda_l}$ conditioned on $(\eta_k)_{k \not\in \Lambda_l}$ be of the form $|\det B_l|\, k_l(\eta)= F(B_l \, \eta^l)$, where $f\in W^{1,1}$ is a probability density, $F_l(\omega)=\prod_{k \in \Lambda_l} f  ( \omega_k) $ and  $B_l:= B_l\big ((\eta_k)_{k \not\in \Lambda_l}\big)$ is a sequence of invertible linear transformations with $\sup \{\|B_l\| \, | \, (\eta_k)_{k \not\in \Lambda_l}\}  \le const\, l^q$. Then  for each $E \in \RR$ exists a constant $C=C(E)$ such that
\begin{equation}
\label{dep}
\EE\l [ \Tr P_\eta^l( \l [E -\epsilon,E \r ]) \r ]  
\le C \  \epsilon \, l^{q+d} , \quad \forall \, \epsilon \ge 0.
\end{equation} 
\end{prp}
Wegner estimates for alloy type models with dependent coupling constants have been established in \cite{CombesHM-1997} (cf.~also \cite{HupferLMW-2001}). The criteria for their validity are formulated in terms of the conditional densities $h_j(\eta)$ of the random variable $\eta_j$ w.r.t.~the remaining ones $\eta^{\bot j}=(\eta_k)_{k\in\ZZ^d\setminus j}$. One necessary condition is that the supremum $\|h_j\|_\infty$ is finite. This motivates the comparison of some properties of the common and the conditional densities of the random coupling constants in the next subsection.

\subsection{Common and conditional densities}
In this subsection we discuss some properties of the common and  conditional densities of the random coupling constants. The proofs of Theorem \ref{theorem1} and Proposition \ref{theorem2} rely on the use of a common density of the type (\ref{k}). Since one has an explicit formula for the corresponding conditional densities, it is desirable to analyze their properties. Particularly, it is of interest whether the conditional densities are bounded by a constant. If this is the case, the Wegner estimates of \cite{CombesHM-1997} and \cite{HupferLMW-2001}  apply by considering the indefinite potential $V_\omega$ in its representation $V_{B\eta}$ as an alloy type potential with dependent coupling constants. 

Lets look first at the common density $k_\Lambda$. Its supremum is easily seen to be $|\det B_\Lambda |\, \|f\|_\infty^L$, so it diverges exponentially with the volume of the cube $\Lambda$. However, this does not matter since in (\ref{sup}) one takes the integral instead of the supremum. 

We restrict ourselves now to the one dimensional $d=1$ case and note that $\Lambda=[0,l[$ and $\Lp= \{-1,\dots,l-1\}$.

Instead of considering the conditional density $h_j(\eta)$ of $\eta_j$ w.r.t~{\em all} coupling constants we will  study (for two simple examples) the conditional density $\rho_j(\eta)=\rho_j^l(\eta)$ of the variable $\eta_j$ with respect to the remaining coupling constants  $\eta^{\bot j}=(\eta_k)_{k\in  \Lp\setminus j}$ in $\Lambda^+$. It is given by $\rho_j(\eta)=\frac{k(\eta)}{g_j(\eta)}$. Here $g_j(\eta):=g_j^l(\eta)= \int k(\eta) d\eta_j$ denotes the marginal density. 
The question is whether $\sup_j\rho(\eta)$ is finite. If it is finite, does the upper bound depend on $l$ and  does the bound diverge as $l$ tends to infinity?

\begin{bsp} 
\label{bsp}
Let $u=  \chi_{[0,1]}-\chi_{[1,2]}$, i.e. the convolution vector be $(\alpha_0 , \alpha_1) =(1,-1)$. Then the assumptions of Theorem \ref{theorem1} are (just) not satisfied, since $|\alpha_1|=|\alpha_0|$. However, $A_l=A_{\Lambda}$ has an inverse $B_l$ with entries $b_{j,k}=\chi_{\{j\ge k \}}$ for all $-1\le j,k\le l-1$. Thus $\|B_l\|=l+1$ and  (\ref{main-estimate}) implies  for $f\in W^{1,1}$
\begin{equation}
\label{a1=1}
\EE \l [ \langle \phi , \chi_j P_\omega^l(I) \chi_j  \phi \rangle \r ] \le 
|I| \ \| f' \|_{L^1} (l+1).
\end{equation}
This gives  a Wegner estimate like (\ref{resultat1}), but with $l^d$ replaced by $l^{2d}=l^2$.

Specialize  now  to the density $f(x)=-4|x|+2$ on $[-1/2,1/2]$ and zero elsewhere. We calculate a lower bound on the supremum by evaluating the conditional density at $\eta=0$, the unique maximum of $k(\eta)$. For $j\in \{-1,\dots,l-1 \}$ and $l-j$  even we have
\begin{equation}
\label{lin}
\sup_\eta \rho_j^l(\eta) \ge \rho_j^l(0) =l-j+1  .
\end{equation}
This bound diverges as the length $l$ of the interval tends to infinity. Thus the mere invertibility of $A$ is not sufficient to ensure the uniform  boundedness of $\sup_\eta \rho_j^l(\eta)$.
Remarkably the bound (\ref{lin}) has the same volume growth as the one in  (\ref{a1=1}) involving the common density. If (\ref{lin}) is actually the exact behaviour of $\| \rho_j^l\|_\infty$ then the conditional density may be used to derive (\ref{a1=1}) and a Wegner estimate. 
\end{bsp}

\begin{bsp} 
Consider now the density function $f=\chi_{[0,1]}$ and the single site potential $u=  \chi_{[0,1]}-\alpha \chi_{[1,2]}$ with $-\alpha=\alpha_1\in]-1,0[$. To the corresponding alloy type model the Wegner estimate in Proposition \ref{theorem2} applies. We calculate now the supremum of the conditional densities $\rho_j$. The common density is given by $ k(\eta) = \prod_{k=-1}^{l-1} \chi_{[0,1]} (\sum_{\nu=-1}^k \alpha^{k-\nu} \eta_\nu)$. For 
$
\eta_{j+1}\in [0,1], \eta_k =0, \forall k\not= j+1$ we have $k(\eta)=
\prod_{k=j+1}^{l-1} \chi_{[0,1]} (\alpha^{k-j-1} \eta_\nu)=1.
$   
 The marginal density 
\begin{eqnarray*}
g_j(\eta) 
&=& \prod_{k=-1}^{j-1} \chi_{[0,1]} \left (\sum_{\nu=-1}^k \alpha^{k-\nu} \eta_\nu \right)\, \int
\prod_{k=j}^{l-1} \chi_{[0,1]} \left(\sum_{\nu=-1}^k \alpha^{k-\nu} \eta_\nu\right) \, d\eta_j
\\
&\le&
\int  d\eta_j \, \prod_{k=j}^{j+1} \chi_{[0,1]} \l (\sum_{\nu=-1}^k \alpha^{k-\nu} \eta_\nu \r) 
\end{eqnarray*}
has for $\eta_{j+1}\in [0,1], \eta_k =0, \forall k\not\in \{j, j+1\}$ the upper bound $ \int_0^1 \chi_{[0,1]} (\alpha \eta_j +\eta_{j+1})d\eta_j \le \alpha^{-1} (1-\eta_{j+1})$. Particularly, $g_j(\eta) \searrow 0$  for $\eta_{j+1}\nearrow 1$ and thus
\begin{eqnarray*}
\|\rho_j\|_\infty =\infty.
\end{eqnarray*}
So proofs of a Wegner estimate which require the conditional density to be bounded cannot be applied to this alloy type potential.
\end{bsp}

Note that we discussed only mathematical aspects of the use of the common, resp.~conditional density. A different question is whether for physical models  the dependence between the coupling constants is most naturally expressed in terms of the conditional or common densities, in terms of the correlation coefficients or more general mixed moments.

\subsubsection*{Acknowledgements}
It is a pleasure to thank T.~Hupfer, W.~Kirsch, V.~Kos\-try\-kin, K.~Veseli\'c and S.~Warzel for stimulating discussions and helpful comments. This work was partially supported by SFB 237:~``Unordnung und gro\ss e Fluktuationen'', the Ruth-und-Gert-Massenberg-Stiftung, both Germany, and the Croatian Ministry of Science.

\def\cprime{$'$} \def\cprime{$'$}

\end{document}